%% file: main.tex
\documentclass{article}

\usepackage{arxiv}

\usepackage{cite}
\usepackage{graphicx}
\usepackage{hyperref}
\usepackage{soul}
\usepackage[framemethod=TikZ]{mdframed}
\usepackage{dblfloatfix}
\usepackage{tabularx}
\usepackage{xcolor, colortbl}
\usepackage{multirow}

\title{Interspecies information systems}

\date{\today}

\author{ \href{https://orcid.org/0000-0002-8597-3156}{\includegraphics[scale=0.06]{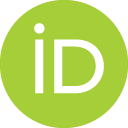}\hspace{1mm}Dirk~van~der~Linden}
\\
	Northumbria University\\
	\texttt{dirk.vanderlinden@northumbria.ac.uk} \\
}

\renewcommand{\headeright}{forthcoming in \emph{Requirements Engineering}}
\renewcommand{\undertitle}{forthcoming in \emph{Requirements Engineering}}

\begin{document}
\maketitle

\begin{abstract}
This article introduces a new class of socio-technical systems, interspecies information systems (IIS) by describing several examples of these systems emerging through the use of commercially available data-driven animal-centered technology.
When animal-centered technology, such as pet wearables, cow health monitoring, or even wildlife drones captures animal data and inform humans of actions to take towards animals, interspecies information systems emerge.
I discuss the importance of understanding them as information systems rather than isolated technology or technology-mediated interactions, and propose a conceptual model capturing the key components and information flow of a general interspecies information system. I conclude by proposing multiple practical challenges that are faced in the successful design, engineering and use of any interspecies information systems where animal data informs human actions.
\end{abstract}


\input{body}

\section*{Acknowledgment}
I am indebted to Anna Zamansky, Brittany I. Davidson, Georgios Plataniotis, Irit Hadar, Ola Michalec, Sybren de Kinderen, and the anonymous reviewers for their constructive criticism and support.

\bibliographystyle{plain}
\bibliography{refs.bib}

\end{document}

%% file: body.tex
\section{Introduction}\label{sec:introduction}
An information system (IS) is a socio-technical system~\cite{luna2005information} that encompasses people, technology, information, and actions people take based on that information~\cite{hanseth2004guest}. Looking at the system as a whole, rather than just individual parts allows us to reason about systemic impacts of its structure, whether that is of its technology, people, or the informed actions they take---also when those impacts transcend beyond the original system's boundaries.

Engineering information systems and accounting for the complex situated networks~\cite{tatnall2005actor} and social contexts~\cite{lamb2003reconceptualizing} of human stakeholders is well researched.
But the recent emergence and increasing popularity~\cite{harrop2016wearable} of data-driven animal-centered technology---think, `FitBit' for your dog, or smart feeding bowls that allow your pets to communicate with you when they are hungry, bring additional complexity as animals\footnote{While humans \emph{are} of course animals, in this article I use `animal' as a natural shorthand for `non-human animal'.} become involved in our design processes. No longer just resources, they become stakeholders and actors in their own right equal to human actors in the same interconnected network of people, animals, and technology{~\cite{westerlaken2016}}.

Information systems that include animals have so far typically only seen them as unintentional, emergent stakeholders~\cite{tryggestad2013project} or even just as resources, such as farm~\cite{sorensen2010conceptual,koksal2019architecture} or veterinary~\cite{morris1991information} information systems.
In contrast, this newly emerging type of animal-centered technology enables humans to better understand animals, opening interspecies communication channels otherwise left implicit, or worse, misunderstood~\cite{tami2009description,lakestani2014interpretation}.
I argue that by doing so, they give rise to an \emph{interspecies information system} (IIS) where both human and animal are actor and stakeholder.

Requirements engineering (RE) research has emphasized the need to recognize unprecedented levels of design complexity{~\cite{jarke2011brave}}. As a field, RE is no stranger to considering the complexity of designing systems that involve animals. Nearly a decade ago, Mascolo et al.{~\cite{mascolo2011ubiquitous}} set out the challenges of developing wildlife monitoring technology that require in-depth understanding of how to design for animals to the RE community, and more RE research traced its inspiration to our co-existence and interaction with animals{~\cite{gotel2011out}}.
Since then, more RE work has considered the challenge of eliciting requirements from non-humans(cf.\, {~\cite{zamansky2017pushing},\,~\cite{ruge2018requirements}},\,{~\cite{hansen2019}}) and integrating them into development methods{~\cite{van2017agile}}.
Yet, the design and use of an IIS may lead to unexpected real-world challenges that transcend beyond the individual people, animals, and technology they envelop, as existing research investigating the use of animal-centered technology has hinted at~\cite{dirkaci,zamansky2019log,dirkagritech}.
Dog activity trackers may indirectly capture unrelated bystanders' behavior, impacting on privacy, while industrial technology for farm animals may unintentionally reveal commercially sensitive information, and poorly designed wildlife technology may affect more ecosystems than envisioned and intended.

Rather than focus solely on the individual people, animals, and technology, understanding the data these systems capture and process into actionable information which informs concrete behavior, is necessary to anticipate such challenges. For example, data-driven suggestions for animal care may be only based on incomplete or inaccurate data, or be misinterpreted.
An IS point of view on the situated technology will aid in understanding and anticipating both the data flow throughout the system and what challenges may arise from it~\cite{misqdef,baskerville2002information}.
Adopting such a point of view will allow us to consider the context of the system level behavior, and how that behavior emerges through interactions of its individual parts{~\cite{jackson2015requirements}} (i.e., the humans, animals and technologies that make up an IIS), giving us a better chance of determining whether the system can satisfy its requirements{~\cite{jackson2014topsy}}. Moreover, this a point of view will further help to focus on a core challenge in RE research: understanding how we, researchers and practitioners alike, can act as facilitators for broader discussions on the far-reaching effects that these new kinds of systems have on our society{~\cite{duboc2020requirements}}.

Thus, to enable more thoughtful engineering of these interspecies information systems, this article serves as a manifesto calling for more understanding and research into IIS, by:
\begin{itemize}
\item \textbf{Constructing a conceptual model for IIS}, elaborating on their key components, and showing how data-driven interspecies interventions are key to understanding impact; and
\item \textbf{proposing a set of ongoing challenges} specific to IIS following from the conceptual model and the flow of information through IIS.
\end{itemize}

The rest of this article is structured as follows. Section~\ref{sec:background} discusses how and why research on animal-centered technology hitherto lacks an information systems perspective. Section~\ref{sec:analyis} derives an initial conceptualization of what IIS are by analyzing `what's on the market', followed by a conceptualization of a model for the IIS in Section~\ref{sec:model}. I set out ongoing challenges that emerge from the information flow in IIS in Section~\ref{sec:challenges} and conclude with opportunities for research and practice in Section~\ref{sec:conclusion}.

\section{Background \& Motivation---the need for an information systems perspective}\label{sec:background}
Human development of technology used \emph{with} animals is not new in human history. Already thousands of years ago, farming tools such as plows were used, first by humans, then with animals to improve conditions for planting crops and thereby increase our yields. More recently, digital technology is increasingly designed and adopted that is also \emph{for} animals~\cite{harrop2016wearable}---it has become animal-centered. While often driven by commercial interests to e.g., increase farm animal productivity, such technology intentionally contributes to the well-being and welfare of individual animals.
The market for such technology has risen greatly. In smart farming, the proliferation of increasingly diverse biosensor technology for farm animals to support animal health (and through doing so, productivity) has significant economic consequences~\cite{neethirajan2017recent,iwasaki2019iot} all but requiring farmers to adopt such technology. Digital technology for pets is equally seen as particularly attractive to investors~\cite{Waters261}. Research has noted how increased economic prosperity has brought with it increased pet ownership, and consumer attitudes have paved the way for big-data driven technology which aid in a consumer demand for health and well-being of their pets~\cite{Waters261}. Adoption of such digital technology for pets can already be observed as pets increasingly consume larger shares of household energy~\cite{strengers2016curious}.
Digital technologies used to monitor wildlife have equally grown in scope and application, being used to inform conservation efforts and ecosystem management{~\cite{nguyen2017animal}}. The technologies used for them similarly have grown in sophistication, from from wildlife cameras to more advanced biological sensor tags{~\cite{wilson2015utility}} and even and unmanned aerial vehicles{~\cite{hodgson2016precision}}.

Understandably, research has increasingly turned its focus on how to best design such technology and understand their impact and is equally diverse in the kinds of animals it investigates. 
Requirements engineering research has discussed the difficulty yet importance of eliciting (and understanding!) animal stakeholders' requirements~\cite{zamansky2017pushing,ruge2018requirements,hansen2019} and the need to integrate them into the way we develop technology~\cite{van2017agile}---albeit in an abstract context.
Research has designed interactions to enrich farm animals' physical and mental well-being~\cite{french2017farmjam}, as well as adapted existing digital technology for captive zoo animals to increase public engagement and understanding~\cite{webber2017kinecting} by allowing zoo visitors to observe animals interacting with technology familiar to them. Extensive work has also been done on the design of digital technology for working animals~\cite{byrne2018dogs,jackson2013fido}, and the adaption of existing digital technology and algorithms to increase the relationship between humans and their pets~\cite{lemasson2013increasing,pons2015envisioning,richardson2017careful,ladha2013dog}. As a result, research on animal-centered technology has built an extensive understanding of how technology might serve animals and the people they interact with.

An important development in this context is the emergence of the field of Animal Studies (AS), a growing interdisciplinary field that seeks to
understand how humans study, conceive of, and interact with animals{~\cite{waldau2013animal}} by taking a critical approach to how we educate ourselves and others about our relations with animals{~\cite{waldau2016second}}.
AS researchers continue to critically investigate the role that technology plays in how humans perceive animals and interact with them.
For example, AS research has argued that dairy farming technology, while ostensibly for animal welfare benefits, is designed in a context where the dairy cow has been shaped by humans to our benefit{~\cite{wicks2018demystifying}}. Indeed, as some other researchers argue, a focus on improving animal welfare in farming (whether or not through technological means) may simply be a solution to deal with the cognitive dissonance of consuming animal products{~\cite{leroy2017animal}}. Other research has argued how wildlife monitoring technologies, while having the potential to improve species companionship, lead to further reinforcement of human and technological dominion over animals due to its asymmetric nature{~\cite{kamphof2013linking}}. Of particular relevance to IIS, AS researchers have critically analyzed new Internet of Things (IoT) technologies and argued they may cause harm to animals, in no small part due to their \emph{anthropocentric} nature of pursuing primarily human interests{~\cite{evans2019there}}.

A second important development in this context is the emergence of the field of animal-computer interaction (ACI)~\cite{mancini2011animal}---tasked with countering exactly that anthropocentric narrative and nature of the design of technologies intended for animals. As a field, ACI was derived primarily from human-computer interaction (HCI), which has promoted critical reflection on the way animals interact with digital technology. Much of the research focusing on this topic has subsequently approached the topic from a background of interaction design, Whether proposing theories for human-animal interaction design~\cite{westerlaken2013digitally,metcalfe2015multispecies,veselova2019implications} or advocating for the application of user-centered design from the animal's point of view to the technology per se~\cite{north2016frameworks,pons2017towards,french2017exploring,ritvo2014challenges,ruge2019method}.
Indeed, a report on research methods employed within ACI~\cite{zamansky2017report} noted most methods employed within the field are effectively borrowed from interaction design, proposing extensions primarily borrowing other disciplinary methods to further ground interactions of animals with technology. A review of ACI research seven years after the publication of its manifesto similarly concluded its research has been largely born out of HCI and ethology, focused on interaction perspectives, calling for more synthesis with animal cognition and behavior~\cite{hirskyj2018seven}.

Notably, ACI research has tended to focus on designing these interactions and technology from the ground up, sometimes eschewing technology already on the market in favor of novel prototypes, whether by researching usable interfaces for cat location trackers using research prototypes rather than commercially available and used technology~\cite{swagerman2018visualizing}, or the speculative analysis of the impact of such technology through the integration of fictional animal personas as stakeholders~\cite{frawley2014animal}.

Yet, as popularity of animal-centered technology has soared over the recent years and consumers and industry alike increasingly use commercially available devices~\cite{harrop2016wearable}, \emph{there is now an ever increasing urgency to study how such technology is actually used and what impact they have on animals and their owners alike}.
Some research has done so, from investigating consumer motivations and barriers to the purchase of companion animal technology~\cite{ramokapane2019does}, to studies comparing motivations for specific types of technology like dog activity trackers~\cite{zamansky2019log}, or contrasting consumer perceptions between technology with similar functionality for human and animal use~\cite{van2020pets}. Going beyond the interaction perspective, little research has considered the use of commercially available technology for working animals and companion animals to understand the importance of \emph{the data such technology capture and process}. For instance, by showing differences in volunteer and organizational apprehension to using activity trackers for blind guide dog raising based on a fear of data protection compliance~\cite{zamansky2018activity}, to showing privacy concerns for pet location data is related to the kind of pet and strength of the bond between owner and pet~\cite{van2019not}, to how technology-supported dog parks could encourage community connections and animal behavioral awareness~\cite{kresnye2019barks}

While the research discussed above indeed investigates the actual technology used on the market, it still lacks critical analysis and evaluation of how such technology is situated in a wider socio-technical context, which is a crucial for engineering an information system~\cite{baxter2011socio,chen2010information,teubner2013information,wieringa2014design}. As a result, it is currently difficult to assess how such technology actually informs understanding of animal behavior and in turn steers human action.

Thus, I propose the need for interspecies information systems (IIS), moving beyond the limited focus of investigating just the technology itself or how human and animal actors interact with them, and engaging in a more holistic analysis of how such technology give rise to an IS that steers and drives human behavior towards animals.

A key requirement for such analysis needs to be its practical focus. As a major recent study among RE practitioners{~\cite{franch2017practitioners}}{~\cite{franch2020practitioners}} has shown, one underlying factor of negative views from practice towards RE research is its perceived focus on challenging research topics, rather than \emph{practical challenges}. The effort to study IIS thus needs to be necessarily driven by \emph{practical} challenges, clearly identifying what IIS are, how they inform behavior, and what challenges arise during their design, to directly benefit design thinking in practice.

IIS, as a specialized field of study, thus ought to investigate the IS that emerge through the use of data-driven animal-centered technology and what practical challenges are faced in their design by focusing on the:
\begin{enumerate}
    \item \emph{actual technology used on the market}, investigating how and with what goal consumers use them;
    \item \emph{way they inform and steer behavior}, investigating how animal data capturing and processing increases human understanding and leads to concrete behavior;
    \item \emph{impacts of these behaviors}, by systematically understanding their impact on human and animal actors.
\end{enumerate}

\section{Understanding what interspecies information systems are}\label{sec:analyis}

\subsection{Examples from real-world use}\label{sec:examples}
The data-driven animal technology focusing on measuring and \emph{suggesting} interventions that are on the market are primarily focused on domesticated animals, that is, animals with whom we as humans have a mutual relationship affecting caregiving and reproduction. This includes distinct categories like companion animals, such as dogs (\textit{canis familiaris}) and cats (\textit{felis catus}), and farm animals such as cows (\textit{bos taurus}). The focus of such technology on such domesticated animals likely reflects our closer relationship to these species and the need for support  in our interspecies caregiving.

These examples are based on extensive research and interactions with vendors of animal-centered technology, market analysis reports, and insights from recent research which has analyzed over 8000 Amazon reviews of commercially available animal-centered technology (activity trackers, location trackers, etc.)~\cite{van2020pets}, and studies investigating users of commercially available animal wearables~\cite{zamansky2019log} and their perceived impact~\cite{dirkaci}.

\subsubsection{Companion animals}\label{sec:companionanimalsexamples}
Companion animal wearables are a quickly growing sector in the companion animal industry and cover a variety of, often data-driven, technology. While such technology is primarily visible in the context of pets, they may be suitable for a given species regardless of it's exact role. That is, technology designed for dogs may be equally useful and suited for pet dogs, working dogs (e.g., detection dog, search and rescue dog), or service dogs (e.g., blind guide dogs or emotional assistance dog). Pet wearables in particular have been noted as one of the top industries for aspiring entrepreneurs to enter, given a large and growing customer base, relatively low investment for entry, and fairly low competition~\cite{bestindustries}. It should be no surprise that there is a proliferation of different data-intensive animal-centered technology being released and promised. Indeed, a recent article in the New York Times~\cite{nytimespets} focused on AI-driven technology for pets discussed the diversity of technology under development and on the market for pets.

Not all of such technology will promote interaction between animals and humans and give rise to an IIS. For example, in 2016, Wagg Pet Foods produced a prototype of a television remote control~\cite{designweek} optimized for dog physiology (color schemes fitting to dog vision, buttons suited to dog physiology). This is an example of a technology developed for use by one species, but not capturing or sharing data with the dog's human owner to inform their understanding of, say, the dog's likes or dislikes for particular TV channels based on their interactions with the remote. It thus remains an isolated technology, rather than giving rise to an IIS.

More potential for an IIS to emerge comes from the growing number of interactive speakers and cameras developed to increase interaction between owners and pets when pets are left alone at home. An example of such devices is Petcube~\cite{petcube}, marketed as an interactive assistant for `pet parents'. Research has shown that the core functionality of such devices makes sense, as pets are capable of interacting with their owners through such technology, such as e.g., dogs using Skype to communicate with their owners~\cite{rossi2016dog}. These devices, while not seemingly giving rise to an IIS yet, make an important first step by \emph{enabling} interspecies interactions.

When such technology goes further, and captures and processes data to inform humans how to structure those interactions, and in doing so \emph{inform} interspecies interactions, they become an IIS. A technology closely related to the interactive assistant shows just such an example. Smart food dispensers are based on such interactive assistants, representing more complex technology where a pet owner can see the amount of food currently in the bowl through a weight sensor, or receive a `communication' from their dog, and in return instruct the technology to dispense food. Such technology may even aid in veterinary care by providing veterinarians with more objective diet information than owner reporting. This shows that such information flow may be both indirect and direct, either when a dog indirectly triggers the system to send a signal to its owner by emptying a bowl of its food, or by doing so directly by barking into the speaker, whereupon the owner may be stimulated to release food. If such interaction is intentional, interspecies communication is indeed enabled by the technology.

Consider one of the more prevalent types of technology for companion animals, pet wearables. The market is filled with devices to monitor location of pets, track their activity and fitness, or even provide detailed insights into their health. These wearables, similar to human wearables, typically exist of a piece of sensor-laden hardware, worn by the pet, and relevant controlling technology, usually in the form of an app for the owner's smartphone. In the context of pet wearables, location trackers are typically based on GPS or RF-based solutions, while activity trackers are typically based on accelerometers, using Bluetooth or WiFi connectivity to share data with controlling devices~\cite{van2019buddy}. A popular activity tracker such as FitBark~\cite{fitbark}, thus consists of a device worn by the dog, measuring its activity akin to a regular human wearable, whose data is processed into human readable form and accompanied by suggestions for interactions (e.g., walk the dog more). Information thus flows from raw captured accelerometer data, to processed human-readable descriptive data of daily activity, to normative instructions informing concrete interactions.

\begin{figure}[htb!]
\centering
\includegraphics[width=.8\columnwidth]{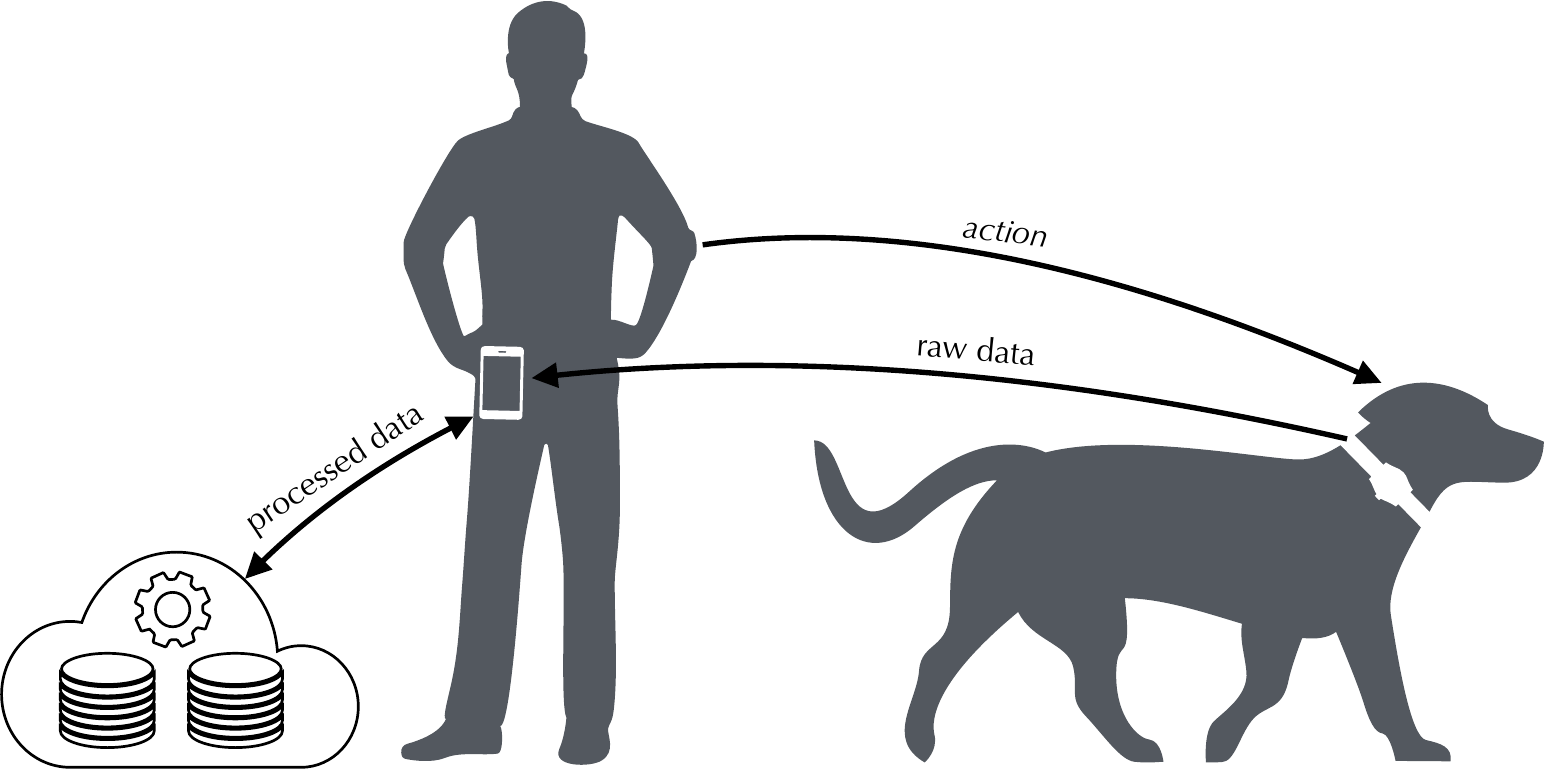}
\caption{How use of pet wearables gives rise to an IIS: hardware worn by a dog sending data to a cloud via the owner's phone, where software converts it into descriptive information and suggestions (e.g., walk more, eat less}
\label{fig:petwearableiis}
\end{figure}

It is this exact information flow that gives rise to the IIS. This therefore goes beyond simple technology enabled interactions, as the IIS provides an information loop in which a data-driven system tells the human owner how to intervene in different aspects of their dog's life, or, more accurately: suggests interventions to different processes affecting their caregiving to the dog. For example, increasing or decreasing activity, increasing or decreasing food and calories based on that activity, and so on. An IIS has emerged, as visualized by the simplified data flow in Fig.~\ref{fig:petwearableiis} consisting of actors of different species, technology capturing data of one species, and processing it for consumption and acting upon by another species. 

\subsubsection{Farm animals}\label{sec:farmanimalsexamples}
Even before technology for companion animals became widespread, farm animals had been subject to increasing use of technology to optimize different processes. 

Similar technology is available for farm animals, often at larger scales. Rather than individual food dispensers, auto-feeding solutions for livestock consist of sensor-driven systems which estimate the amount of food needed for farm animals based on their physiology and environmental conditions, such as e.g., determining feed during developmental stage when growing chicks for poultry meat. Such technology, however, do not give rise to an IIS as they are one-sided data-driven systems which automate the decision-making, taking the human out of the loop to decrease workload. 

Yet, different systems built to optimize scaling and reduction of workload place the human central in the loop. For example, a sensor-based system for the monitoring of health and welfare data of dairy cows. Monitoring technology worn by each individual cow contains sensors which capture activity data and vital signs, sending this towards a central IS where it is processed and visualized for a farm operator to keep track of the physical and mental state of each cow. Based on data analysis, the software can inform the farm operator of cows which are showing indicators of factors that may impact the quality of their produced milk (e.g., stress levels, lameness, overheating), and, just as with companion animal systems we discussed before, \emph{inform} them of concrete interactions that are required to correct this. In this context, this may be both one-on-one interactions between a human and an animal, such as stimulating a cow to walk around, or provide them with additional cognitive enrichment, while it may also be indirect technology mediated interactions, such as turning on air-conditioning to reduce overheating.

\begin{figure}[htb!]
\centering
\includegraphics[width=.8\columnwidth]{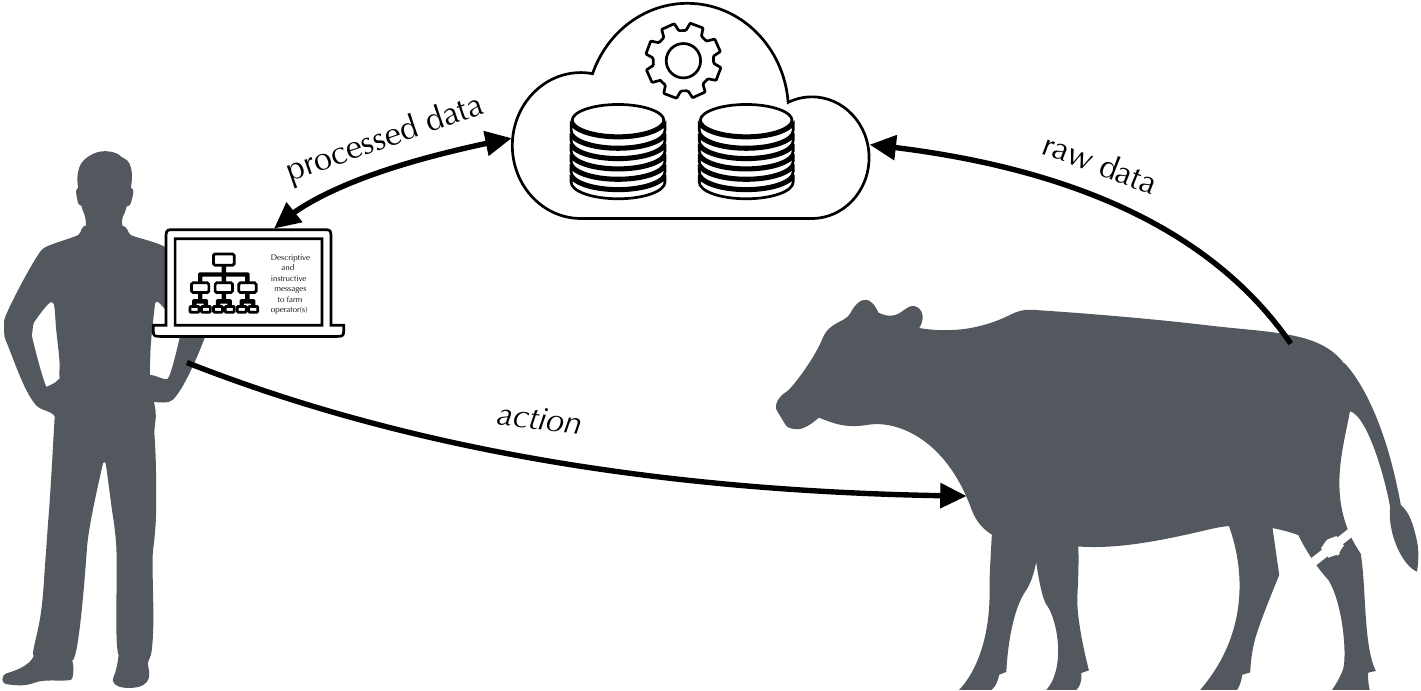}
\caption{How use of sensor-driven systems for farm animals gives rise to an IIS: hardware worn by cows sending data to a central `command points' in the farm, where an operator monitors the physical and cognitive status of each cow, and is instructed to make environmental changes where necessary (e.g., stimulate cows to be active to reduce lying around time, provide cognitive toys, turn on the AC to reduce heat stress)}
\label{fig:farmanimaliis}
\end{figure}

Here, just as with the case of pet wearables, an IIS emerges where a combination of technology measuring data of one species are processed to inform a human actor how to best intervene in the support of a particular process, as visualized in Fig.~\ref{fig:farmanimaliis}.

It would be remiss, however, to move on without explicitly considering that an IIS in the context of farm animals, even if ostensibly meant to inform about animal welfare (albeit with a commercial ulterior motive), does eventually lead to fatal outcomes for many of the animals within the IIS. Consider, for example, the use of data-driven technology to reduce stress for pigs (\emph{Sus scrofa domesticus}) in pre-slaughter phases in order to improve meat quality{~\cite{sardi2020identification}}. An IIS may similarly emerge that processes data (e.g., cortisol level monitoring) to assess and inform whether certain actions should be taken to improve welfare on the short term (e.g., provide cognitive enrichment, avoid particular handling), which is ultimately tasked with the process of improving meat quality on the long-term (cf. the context of using enrichment devices with large pigs for such purposes{~\cite{nannoni2018enrichment}}). Thus, while I phrased that the IIS in this farm context would inform a human actor ``how to best intervene'', it must not be taken for granted that this is in the best interest of animal actors per se---especially when these interventions serve a business process that is not at all concerned with animal best interests (i.e., optimizing meat quality and conditions for slaughter).

\subsubsection{Wildlife}\label{sec:wildlifeexamples}
Wildlife has been subject to human technologies for a significant amount of time. For instance, the use of wildlife crossings, whether tunnels or bridges, that allow wildlife to safely cross human made barriers like highways. Or their inverse, wildlife grids, that discourage animals from crossing into particular areas. These are concrete, physical examples of technologies with an underlying similar purpose---to manage how humans co-exist with wildlife, and to reduce our negative impact on their very existence.

In other words, these are technologies to manage how we co-exist with other species---an increasingly important topic as human civilization encroaches on animal populations and affects the way they survive{~\cite{geffroy2020evolutionary}}. Wildlife monitoring, in particular may be of most interest as a  source of technologies that give rise to an IIS, as these technologies and the data they capture are essential to ``inform conservation and management decisions to maintain diverse, balanced and sustainable ecosystems.''{~\cite{nguyen2017animal}}. Indeed, wildlife monitoring technologies are essential to understand the many ways in which human activity and civilization impact on wildlife{~\cite{wilson2015utility}}. This seemingly less anthropocentric purpose than the companion and farm animal examples I have discussed so far is achieved through a variety of technologies---from simple manual counting to cameras, to even unmanned aerial vehicles{~\cite{hodgson2016precision}}. The data they capture is already raising discussions on what such data contains beyond its primary purpose, and what (malicious) behavior it may unintentionally inform---see e.g., debates on the potential of digital poaching{~\cite{manghi2019}}.

It seems evident, therefore, that wildlife technologies similarly give rise to interspecies information systems (IIS). There is an important distinction that becomes apparent, however, going even beyond the lesser anthropocentric nature of such an IIS. Data collected by these devices, often in major project and group efforts do not tend to inform actions with similar immediacy as in the context of companion and farm animals. Rather, information collected by wildlife technologies (e.g., herd observations, impact of human activity on quality of life) informs separate decision-making processes that set policy towards wildlife, often on governmental level. There is an additional layer of complexity with a decision-making structure that essentially introduces a gulf of execution between the initial observations, interpretations, and the actual actions taken towards wildlife--making these additional layers vital to the actual decision-making that informs the actual actions that then affect wildlife populations{~\cite{linnell2001predators}}.

\begin{figure}[htb!]
\centering
\includegraphics[width=.8\columnwidth]{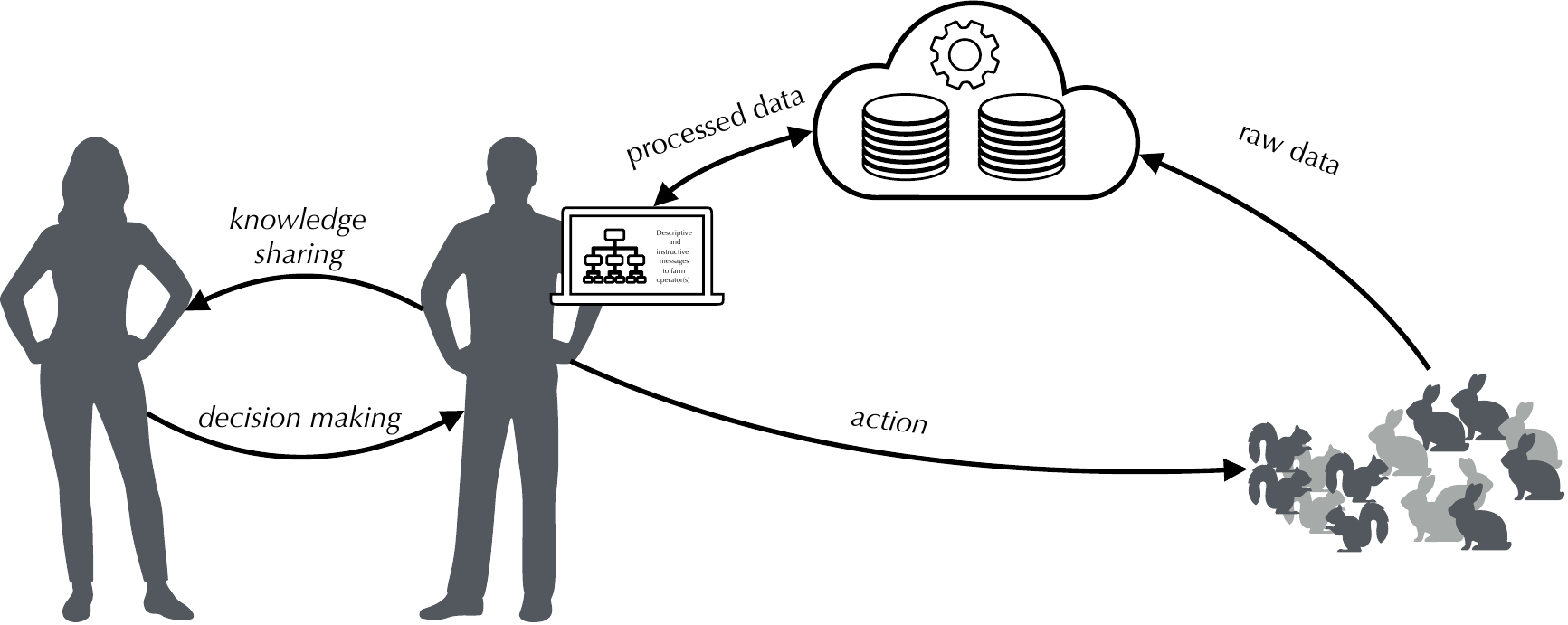}
\caption{How use of wildlife monitoring technology gives rise to an IIS: hardware used to monitor wildlife captures data converted into descriptive information (e.g., population habitats, health, movements), which is then used to inform decision-making regarding wildlife conservation, potentially culminating in policy which then is used to act upon that wildlife.}
\label{fig:wildlifeiis}
\end{figure}

In terms of the IIS that emerges, as Fig.{~\ref{fig:wildlifeiis}} visualizes, the flow of information thus does not go simply from observation to action, but necessarily includes an additional step of interpretation, only then resulting in a mandate for action towards animals.

Given the focus of wildlife monitoring being on the preservation of entire populations of animals, a wildlife IIS may also elicit debate on what it means to ``best intervene'' when animals are in need. While in the case of farm animals human interests likely trump animal interests on the long-term, with wildlife it may be more like that interests of populations of animals likely trump individual animal interests on the long-term. Consider, for example, a wildlife IIS that informs human actors of the need to cull a rapidly growing population of animals to ensure they do not threaten their own survival by entirely depleting food sources. This shows a further nuance in the notion of interventions and impact, in that it may be targeted either at individuals or entire populations of some species.

\subsection{Commonalities}\label{sec:commonalities}
Through these examples, I have shown that the essence of an IIS is to inform humans of action to take towards animals, and established some important commonalities that IIS share regardless of functionality (e.g., activity tracking, health monitoring) or involved species (e.g., companion animals, farm animals, or wildlife). Specifically, that:

\begin{mdframed}[roundcorner=10pt]
\begin{center}\textbf{Commonalities of an IIS}\end{center}
\begin{itemize}
    \item an IIS enables a flow of information across species, typically informing human stakeholders of physiological or behavioral states of another species;
    \item this information informs, often intentionally, informed interventions from one species to another (whether to individual or groups), typically to affect their physiological or behavioral state (whether positively or negatively); and
    \item those interventions impact a species (whether individuals or groups) to aid in an external process (whether informal or well defined).
\end{itemize}
\end{mdframed}

It is therefore important for the informed design and use of IIS that they understand in detail where and how information is created, and how it flows between components of the system.

Below I discuss how these commonalities may be informed by, and grounded in, relevant theoretical frameworks, in turn informing a model of a general IIS.

\subsection{Interspecies information flow}
The flow of information between species in an IIS is typically meant to enable interactions, or communication between distinct species.
From the microbial level to interaction of different mammalian species, interspecies communication has been studied extensively. From differences species of old world monkeys (\textit{cercopithecus}) having mutually intelligible warning calls~\cite{zuberbuhler2000interspecies}, play between chimpanzees (\textit{pan troglodytes}) and bonobos (\textit{pan paniscus})~\cite{lyn2006development}, to the well studied interspecies communication between dogs and humans, showing the formation of such communication even from a young age~\cite{dorey2010domestic}. Research has suggested that we should not restrict ourselves solely to reciprocating communication, as unilateral `interactions' play a major role in maintaining interspecies communities~\cite{mougi2016roles}.
Kostan proposed a theory of interspecies communication~\cite{kostan2002evolution} based on assessment and management of information which provides insight into how the direction of information flow may enable interventions in increasing levels of reciprocity. It classifies interspecies communication into:
\begin{enumerate}
    \item Unidirectional assessment (one species acting upon another species' intra-species communication)
    \item Bidirectional assessment (both species acting upon each other's intra-species communication)
    \item Asymmetric communication (one species informing another species)
    \item Symmetric communication (two species information one another)
\end{enumerate}

Assessment, whether uni- or bi-directional, are not relevant to understanding the information flow in an IIS as these constitute one-sided `consumption' of information, where no interaction between species is enabled. For example, in the context of dogs, an example of unidirectional assessment could be a person hearing several dogs barking loudly in a street, and inferring that it must be a sign of danger, hence deciding to avoid walking down that street. As the enabling of communication is key for IIS to emerge, the primary distinction to make is thus whether such communication is \emph{asymmetric} or \emph{symmetric}. From the examples I have discussed above in Section~\ref{sec:examples}, this shows for example:

\begin{mdframed}[roundcorner=10pt]
\begin{center}\textbf{Directionality in an IIS}\end{center}

\noindent \emph{Asymmetric communication:} When an IIS enables an actor of one species to intervene in the behavior of another species. \\
For example, the monitoring of livestock, where a human operator monitors data of a herd of cows and intervenes where appropriately, while cows are unaware of the monitoring. Similarly, pet wearables present an asymmetric information flow, where a dog is monitored and software suggests how the owner may interact with them or intervene in their behavior, while dogs are unaware or engaging similarly in the IIS. \\

\noindent \emph{Symmetric communication:} When an IIS enables actors of multiple species to intervene in each other's behavior. \\
The examples of interactive assistants connected to feeding systems enable symmetric communication. A dog may share information and request action of their owner (e.g., barking to request food), while the human owner similarly may engage in interactions and request action of the dog through audio/video link.
\end{mdframed}

It is thus important for the informed design and use of an IIS to account for the directionality that its information flow enables.

\subsection{Interspecies interventions}
The information flow within an IIS is meant to inform interventions from one species to another. It is thus important to understand how the key components like actors and technology within the IIS relate to each other to enable such interventions.
As a start for theoretical grounding of how different components of an IIS are needed to enable interventions, consider the SHELL conceptual model that describes the interactions between the four main components of a socio-technical system: Software, Hardware, Environment, and Liveware~\cite{carayon2006human}. Each of these components interacts in a given system, where here, in particular, the interactions of actors to other components are key to understanding how an interspecies intervention is enabled. As the examples in Sec.~\ref{sec:examples} already revealed, depending on the exact technology, actors of a given species may only interact with some of the hardware, and these interactions may be passive or active. This means that we need to explicitly distinguish the interactions that actors have with hardware, software, and each other, showing that interspecies interventions are effectively enabled as three successive interactions:

\begin{mdframed}[roundcorner=10pt]
\begin{center}\textbf{Key interactions in an IIS}\end{center}

\noindent \textit{Actor--hardware} interactions can be \emph{active} or \emph{passive},
For example, with wearable technology, animal actors typically have a passive relation to the hardware, simply being made to wear it. Other technology, such as smart food dispensers show passive interactions between animal actors and the hardware, triggering a signal for more food simply by emptying the bowl. Human actors, however, will typically interact with both the hardware worn or used by animal actors in order to ensure its suitability and appropriate fit (e.g., ensuring the animal actor is not bothered by a wearable), as well as separate hardware used to control and monitor these devices. \\

\noindent \textit{Actor--software} interactions, are the critical aspect enabling an interspecies intervention, as human actors consume information and suggestions how to interact with, or intervene in another animal actor's behavior. \\

\noindent \textit{Actor--Actor} interactions, finally, are both the information-driven interventions that a (typically) human actor takes towards animal actors in the IIS to aid in an external process such as caregiving or quality management, and the human-human actor interactions that may first occur as a prelude to informing those cross-species interactions, as e.g., in the context of wildlife management decision-making.
\end{mdframed}

This emphasizes the importance for designers of having a detailed view on what technology actors of different species interact with, and explicitly distinguishing between \emph{human} and \emph{animal} actors in terms of their interactions to other components of the IIS.

\subsection{Interspecies impact}\label{sec:interspeciesimpact}
Many of the interventions that actors make across species boundaries informed by an IIS will lead to concrete impacts on an animal's physical wellbeing, both on the short and long term.
Over twenty-five years ago, Hirschheim et al.~\cite{hirschheim1995information} already noted that IS design is ``not merely a technical intervention but involves social and ethical dilemmas that affect the human, social and organizational domains.'' Interspecies interventions enabled by an IIS showcase this complexity: both human and animal actors, as well as the wider societal and organizational environments in which both species co-exist are affected by the interventions that the IIS suggests--and as Sec.{~\ref{sec:farmanimalsexamples}} has shown, not always for the better. Moreover, the very design of the IIS may even further strengthen assumptions that are (unintentionally) harmful to animals{~\cite{evans2019there}}.
As interspecies relations are highly complex~\cite{blue2011trans}, it is therefore important to systematically treat the (potential) impact that actors have on actors of other species.

Understanding interspecies interventions enabled by an IIS as symbiotic relationships, allows us to distinguish between different levels of harm and benefit to the involved actors of different species~\cite{douglas2010symbiotic}, including relationships that are:
\begin{enumerate}
    \item Amensalistic (harming one species, while not affecting the other)
    \item Parasitic (benefiting one species, while harming the other)
    \item Commensalistic (benefiting one species, while not harming the other)
    \item Mutualistic (benefiting both species)
\end{enumerate}

The impact of interventions enabled by an IIS through the subsequent interactions described before can thus be described in increasing levels of desirable symbiosis:

\begin{mdframed}[roundcorner=10pt]
\begin{center}\textbf{Impact levels in an IIS}\end{center}

\noindent \emph{Amensalistic} impact, to a certain extent, may primarily arise unintentionally during the design and use of an IIS if technology is designed without due regard for all involved actors. This may be linked to hardware, such as for example wildlife technology leading to unintentional death of its subjects~\cite{cid2013preventing}. Perhaps more critically for an IIS, software-linked harms may arise if interventions suggested by the IIS need are not accurate and appropriate. For example, a dog owner may unintentionally cause musco-skeletal injury in a dog through overtraining as a result of erroneous advice generated by an activity tracker. \\

\noindent \emph{Parasitic} impact is unlikely to arise as an intended consequence of an \emph{individual} interspecies intervention. However, interventions are in the aid of external processes--not all of which will serve the best interest of animals in the long term. We might thus critically assess whether the long-term benefits of processes supported or enabled by these interventions harm a species in the IIS, while benefiting another by e.g., trivializing their caring needs while giving a human owner a false sense of security in their caregiving capability, or more commonly, preparing farm animals for slaughter. \\

\noindent \emph{Commensalistic} impacts are seen in e.g., the examples of farm animal technology used for short-term beneficial interventions. The interventions that the IIS in Fig.~\ref{fig:farmanimaliis} enables give direct relief to the cows by e.g., reducing their heat stress, or providing cognitive enrichment, but do not directly provide benefit to the human actor in the IIS. Rather, as a reverse of parasitic impacts, here benefit may more likely arise on the long-term as a result of the external processes that these interventions support by e.g., increasing the quality of the produced milk, which in turn brings commercial benefits. \\

\noindent \emph{Mutualistic} impacts of an IIS are when an intervention benefits both species of actors. Pet wearables, activity trackers in particular, provide an example of such benefit. A typical concrete intervention that a dog activity tracker may suggest is to simply take the dog for a walk. As research into the motivations and actual use of dog activity trackers has shown, the use of these trackers leads to improved activity and potential health benefits not only for the dog, but to increased motivation for fitness of the human owner as well~\cite{zamansky2019log}.
\end{mdframed}

The critical reader may recall that Sec.{~\ref{sec:wildlifeexamples} discussed another complication beyond that of balancing short and long-term impacts. Indeed, it may be the case that interspecies interventions harm some individuals of a species, or even prove to be fatal. For example, killing (`culling') a set number of animals in a population of wildlife species, benefiting their population as a whole by ensuring the population does not deplete its food sources to the point of no recovery. Here, a seeming amensalistic interspecies intervention (killing the set of number of wildlife animals) leads to an eventual commensalistic outcome (benefitting the wildlife, while not directly benefiting the human actors involved (although, of course, on a much longer term, these actions can be interpreted as benefiting human actors by ensuring stability of our overall shared ecosystems). Designers of an IIS may thus also need to consider how to balance the short-term and long-term impacts of interspecies interventions their technologies inform on, potentially separating them into separate interventions so as to allow for critical reflection on whether some harm gives way to the greater good.}

\subsection{Does an IIS control or inform?}\label{sec:kindofsystem}
We have now established certain commonalities of IIS, in that they enable an interspecies information flow with a given directionality, in turn informing and enabling interactions that lead to interspecies interventions with concrete, real-world impacts on one or more species. An important consideration designers might reflect on is whether that means an IIS \emph{informs} interspecies interventions, or whether it \emph{controls} them.

It might be tempting to consider the IIS as a system that \emph{controls} interspecies interventions, akin to a cyber-physical system (CPS) that extends human capabilities in terms of sensing, decision-making and action{~\cite{griffor2017framework}}, while keeping humans in the loop{~\cite{sowe2016cyber}}. There are indeed similarities to CPS from what I have established as key requirements for an IIS--both are preoccupied with enhancing capabilities{~\cite{loucopoulos2019requirements}}, need to be data-driven, and values-based (cf.{~\cite{czarnecki2018requirements}}), concerned with continuous detection of challenges{~\cite{jin2019re4cps}}, and ensuring safety of more-than-human interactions becomes an important requirement (cf.{~\cite{dede2020safety}}).
In some specific contexts, parts of an IIS may indeed be preoccupied with (semi-)automated control---in particular in the context where animals are effectively seen as resource, such as e.g., automated milking systems which control milking processes while also informing a farmer of further potential actions to take. Designing such systems indeed requires significant considerations on data capture and management to ensure adequate control (cf.{~\cite{costa2016design}~\cite{costa2016modeling}}).

Yet, as the considerations and design challenges from other examples analyzed in Sec.{~\ref{sec:analyis}} show, the key essence of an IIS \emph{as a whole} is to \emph{inform} of interspecies interventions to take, while not actually intending to \emph{control} such actions in (semi-)automated ways. Moreover, as we have seen from just the three examples of different technologies analyzed, the interspecies interventions taken suggested by an IIS pass through multiple layers of additional consideration and complexity before culminating in a realized interspecies intervention. In its simplest form, this might mean whether a dog owner followed up on suggestions provided by the processed data. Indeed, deviation where necessary is important so that, e.g., veterinarians can overrule suggestions from the IIS. In the context of wildlife IIS, Sec.{~\ref{sec:wildlifeexamples}} there is an additional gulf of execution that separates the actual apparatus of the IIS itself and the interventions taken: culling of wildlife is certainly informed by an IIS through its monitoring or recording of animals and suggestion of potential interspecies interventions, but before any such interspecies interventions are taken, they go through external decision-making, policy considerations, likely involving entities and resources entirely separate to the IIS itself.

Thus, designers, in their design thinking, ought to think of an IIS as a system that enables information flow between species to inform of interspecies actions to take--whether improved caregiving, optimization of farming, or tracking of wildlife populations, and focus on that information flow and the interspecies interventions it enables.

\section{Modeling interspecies information systems}\label{sec:model}
Conceptual models are important to ``represent phenomena in some domain'' in order to ``facilitate early detection and correction of system development errors''~\cite{wand2002research}. Rather than focus on underlying deep structures, like ontological reality~\cite{wand1995deep} (additionally complicated with multiple species involved as stakeholders), I will take a pragmatic view to construct a model of an IIS that captures the key components identified in Section~\ref{sec:analyis}. Doing so will allow us to understand how the components interact and exchange data that eventually informs interspecies interventions, and more easily inform high-level design thinking of how the different elements of an IIS interact.

\begin{figure*}[ht!]
\centering
\includegraphics[width=1\textwidth]{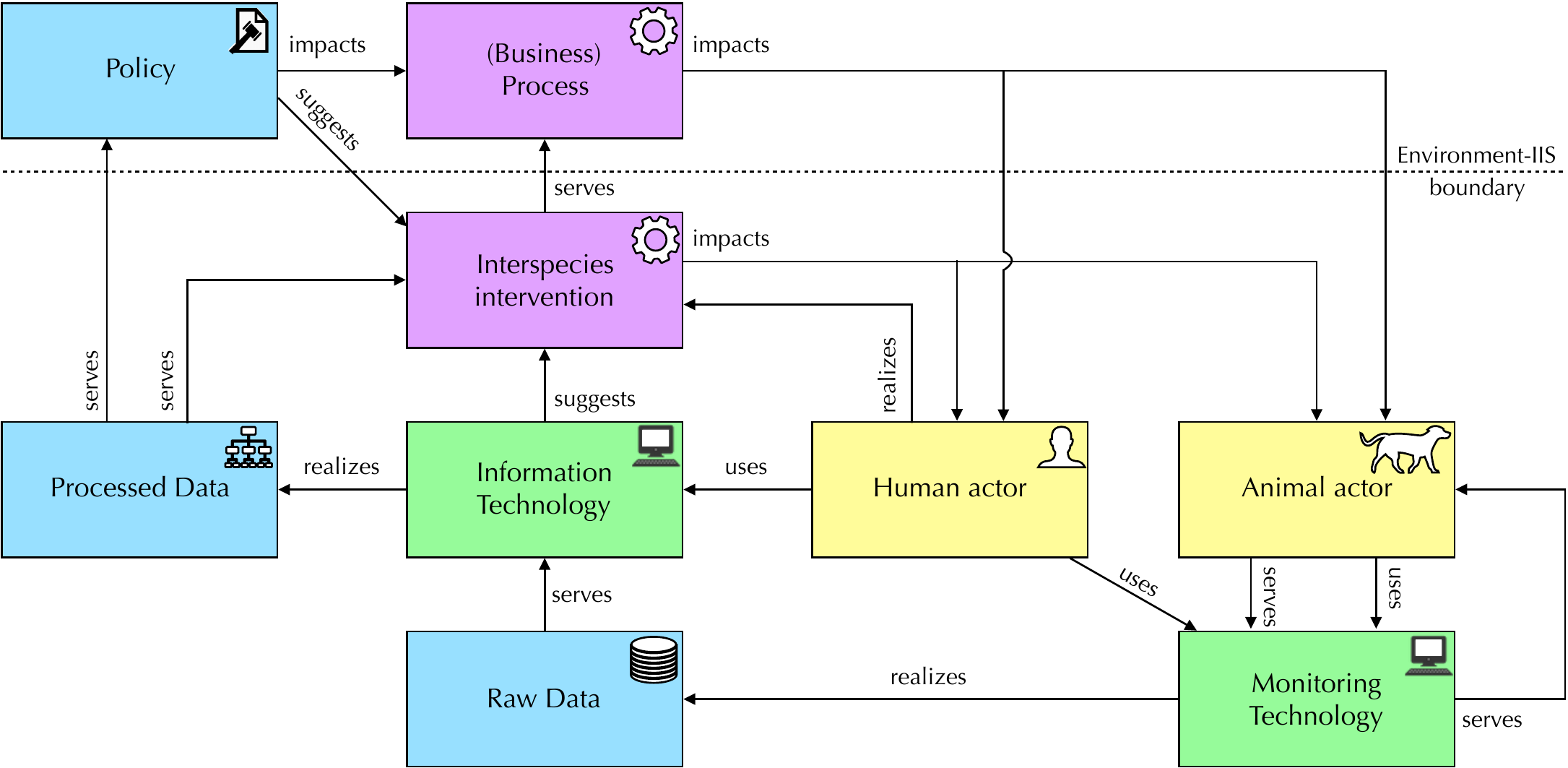}
\caption{Key elements of data-driven interactions within a general IIS.}
\label{fig:metamodel}
\end{figure*}

As it is now established that an IIS is a system where information flow enables interspecies intervention, which in turn have impact on actors of different species, the model visualized in Fig.~\ref{fig:metamodel} reflects these key elements and the relationships between them.

The model shown in Fig.~\ref{fig:metamodel} captures how the interactions between the components of the IIS enable a data flow which informs concrete interspecies interventions and affects processes outside its own boundaries.
Animal actors `serve' as input to the \textsc{monitoring technology} hardware, whether a wearable like dog activity trackers or cow vital sign sensors, or even an interactive assistant, which in turn captures \textsc{raw data}---such as accelerometer data, location, or audio/video recordings.
The \textsc{information technology}, typically consisting of software running on a human's smartphone or desktop computer realizes \textsc{processed data} and suggests \textsc{interspecies interventions} which are in turn realized by a \textsc{human actor}. Alternatively, \textsc{processed data} may first inform decision-making processes external to the IIS, resulting in \textsc{policy} that in turn suggests (refined) \textsc{interspecies interventions}. These interventions both impact upon an external \textsc{process}, such as for example pet caregiving, as well as directly impact the \textsc{animal actor} and/or the \textsc{human actor}.
Putting this in context of the examples analyzed in Section~\ref{sec:examples}, Table~\ref{tab:definitions} gives two more detailed examples of a partial instantiation of an IIS to support a specific process for a companion animal and farm animal scenario.

\input{table1}

\begin{figure*}[htb!]
\centering
\includegraphics[width=1\textwidth]{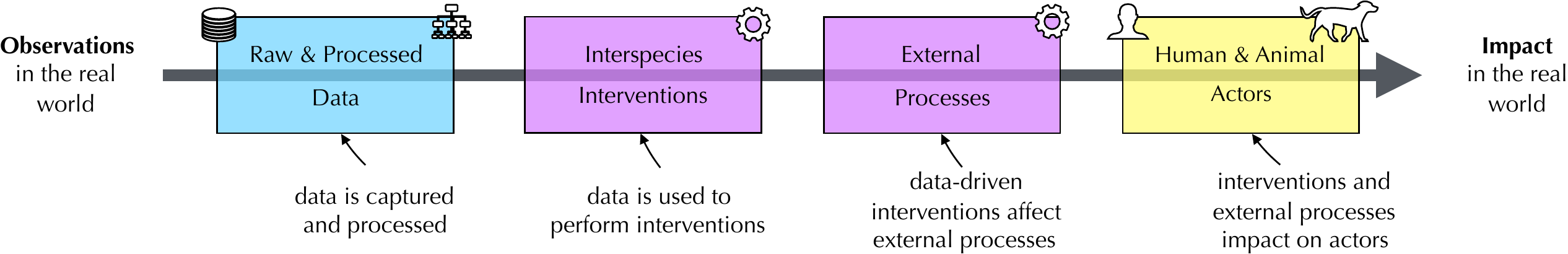}
\caption{Flow of data through key components of the IIS as it transforms observations in the real world to impact.}
\label{fig:flow}
\end{figure*}

Figure~\ref{fig:flow} shows linearly how the flow of data from its initial capture by the monitoring technology, to eventual impact in the real world following a human actor's actions, involves several steps of translation and interpretation. 
Thus, while it is the human actor who performs an interspecies intervention, they do so predicated upon data suggested by the IIS, which goes through several translation and interpretation steps.

This is important to consider, as limitations and biases may creep in at several stages which affect the interventions finally performed:
\begin{enumerate}
    \item First, raw data is captured by monitoring technology with a specific set of sensors, presenting a limited model of of the animal's reality.
    \item Second, processed data is generated by information technology with a specific set of algorithms, presenting a further limited model of reality, and incorporating a potential set of biases by focusing solely on them.
    \item Third, the processed data is then presented in a particular form and suggested to a human actor, who needs to interpret it in order to make an interspecies intervention, presenting a further limitation according to their potential biases and willingness to accept the suggestion.
\end{enumerate}

This poses a challenge to ensuring that the interspecies interventions suggested by the IIS are appropriate, and acted upon appropriately---these and more data-driven challenges I discuss in more detail below in Section~\ref{sec:challenges}.

\section{Ongoing Challenges for engineering interspecies information systems}\label{sec:challenges}
\begin{figure*}[hbt!]
\centering
\includegraphics[width=\textwidth]{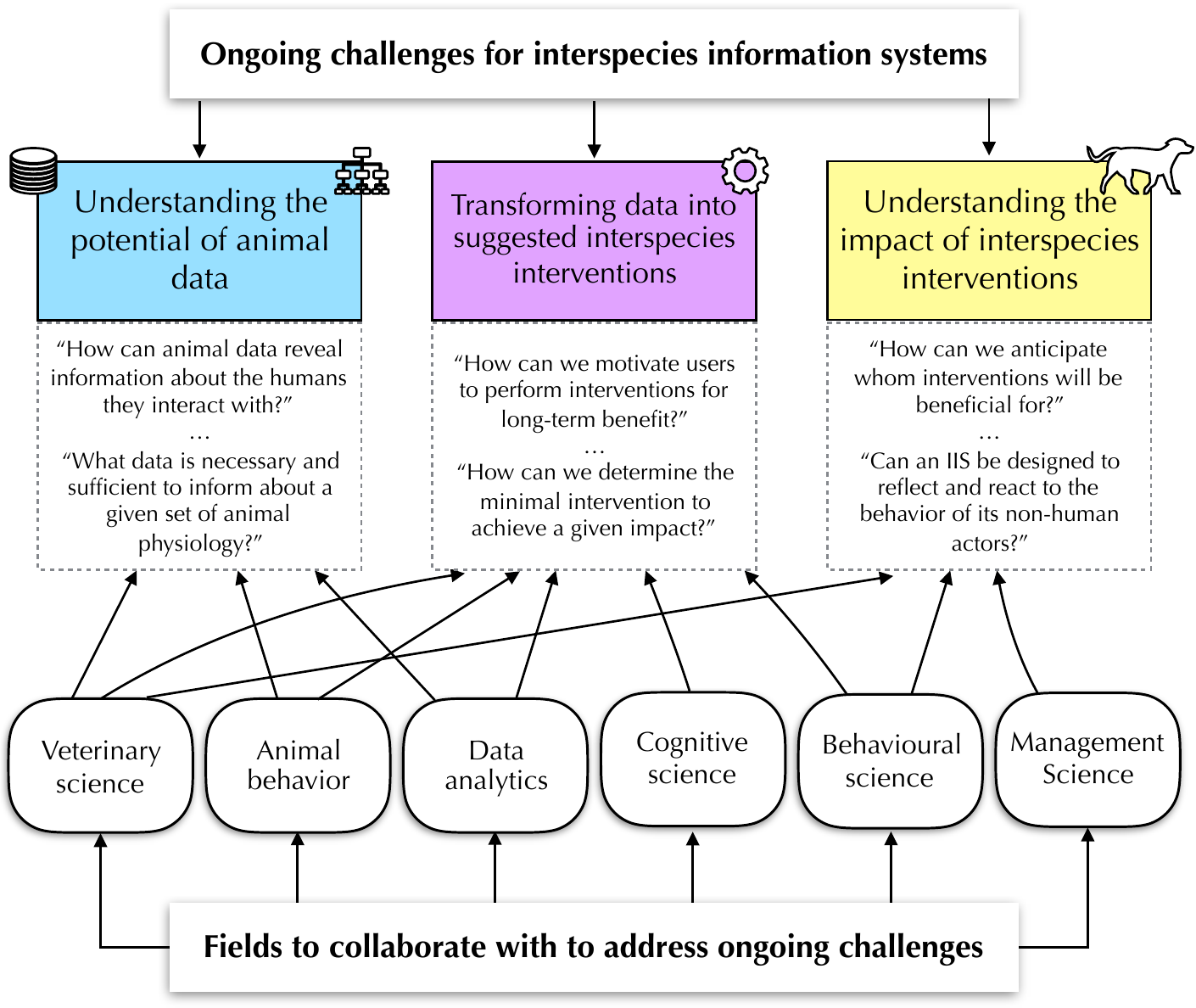}
\caption{Ongoing challenges for IIS with example research questions and the fields of study needing to collaborate to tackle them.}
\label{fig:research}
\end{figure*}

This section will focus on challenges that need to be considered during the engineering of an IIS\footnote{This article does not intend to reinvent the wheel when it comes to IS engineering in general, hence, I focus on challenges specific to the animal data flow that sets IIS apart, acknowledging the importance of understanding general IS success and failure factors~\cite{urbach2009state,yeo2002critical}.} as visually summarized in Fig.~\ref{fig:research}. It sets out what challenges need ongoing considerations during the analysis \& design, implementation, and maintenance of an IIS to ensure its success, and what other expertise is needed to achieve it.
As mentioned in Section{~\ref{sec:background}}, this necessarily focuses on practical challenges that arise during IIS engineering, and what should be done to tackle them. Or, as Davis and Hickey said, RE research should follow the rule of ``know thy customer'' when it comes to ensuring our work is helpful for those we aim to support{~\cite{davis2002requirements}}. In this case, those designers engaged in the design of technologies that may give rise to IIS, and future designers of IIS. Thus, to lower the barrier of technology transfer from RE research to practice{~\cite{kaindl2002requirements}}, I take a decidedly pragmatic approach here describing challenges that \emph{will} be faced in the design and development of new IIS, regardless of whether research considers them solved in theory.

Importantly, these are considered \emph{ongoing} challenges in the sense that they are dependent on the socio-cultural and technical context in which the IIS operates. They are unlikely to be exhaustively solved, as new contexts like changing legislation, technology, insights on data, algorithms, cultures, or shifting attitudes, will continue to bring these challenges to the foreground. Thus, I propose these challenges to guide ongoing research on IIS and invite researchers and practitioners to ensure IIS design and use always accounts for the human and animal stakeholders it serves.

\subsection{Understanding the potential of animal data}
The overarching question framing this challenge is \emph{what is contained in the animal data an IIS captures?}

The immediate value of animal data to the IIS is to assist human actors in improving interspecies communication by allowing humans to understand and respond to animal signals they might not otherwise understand. Like other industries, the practical approach to designing sensor-driven systems for animal understanding has been understandably led by a ``more, more, more'' approach to what kind of data to capture. But with each additional type of data we capture and process in an IIS we need to carefully consider several points:
\begin{enumerate}
\item \emph{What do data capture?} If you want to measure activity, an accelerometer is likely enough, while precise location tracking will need GPS, and in turn health trackers will require more detailed bio-sensors. Determining what phenomena in the real world the IIS is meant to inform about is critical in considering what data and thus sensors are needed. Moreover, accurately classifying that data to objectively reflect the phenomena are important---accelerometer data without adequate classification algorithms brings no value.
\item \emph{What \emph{else} do data capture?} Even if the IIS uses a minimal set of sensors to capture a particular phenomenon, like using an accelerometer to capture activity data, does not mean that more information cannot be gleaned from that data. Careful consideration of what else can be inferred is necessary to understand the value of the data to the IIS, and to other, potentially adversarial, stakeholders.
\end{enumerate}

\subsubsection{What do data capture?}
The goal of an IIS determines what phenomena it should capture and inform about, making the question of what data could inform users about animal behavioral or physiological phenomena the first most important design concern. The technology within the IIS capturing that data, however, needs to still be worn and used by animals, which requires them to be as inobtrusive as possible in order not to interfere with the animal's natural behavior which can lead to unnatural data or worse: harm to the animal or even death~\cite{cid2013preventing}. Therefore, in many cases it is not feasible to `just' incorporate as many sensors as possible to capture as much data as possible. This means trade-offs need to be considered between on the one hand ensuring the design of the monitoring technology is unobtrusive and suits the animals physiology, while still enabling as much data-driven value to the IIS as possible. Finally, as the purpose of IIS is to stimulate actual technology being used on the market, rather than proposed scientific prototypes, simple business considerations need to be taken into account, to strike a balance between e.g., cost of sensors (and similar sensors of increased resolution, such as incorporating 9-axis instead of 3-axis accelerometers), and the additional data resolution they bring.

Moreover, when the hardware has been designed, and it can capture data, there is still work to be done to ensure it accurately reflects the phenomena it is meant to capture~\cite{lawson2016power}. The value of classification algorithms (cf.~\cite{ladha2013dog}) cannot be understated, as the accurately and correctly processed sensor data forms the foundation for any further data processing into human-readable descriptions and suggestions for interspecies interventions. Thus, the design of the hardware and software that captures and processes the sensor data needs to be done in collaboration with those who understand animal physiology and behavior. Besides veterinary science, the emerging field of computational ethology~\cite{anderson2014toward} is fundamental to involve in the design of such algorithms, as it attempts to solve some of the challenges with classical ethology as the study of systematizing animal behavior by automating it to allow for more scaled, faster analysis, and increase objectivity by reducing reliance on human observers.

Some examples of questions that should be raised during the engineering and research into a novel IIS to address these matters thus include:
\begin{itemize}
    \item What data is necessary and sufficient to inform about a given set of animal behavior and/or physiology?
    \item How can we systematically translate desired animal phenomena into sets of sensor (data)?
    \item How can we strike a balance between maximizing captured data on the one hand, and the restrictions imposed by animal physiology and behavior, as well as business reality, on the other hand?
    \item What computational ethology systems can accurately classify sensor data into accurate phenomena descriptions?
    \item What value can longitudinal animal data bring for understanding and predicting situated animal health and behavior?
\end{itemize}
To do so will require IIS researchers to involve not just data analytics, but work together with experts from veterinary science and computational ethology to understand how data reflects real-world animal physiology and behavior.

\subsubsection{What else do data capture?}
It is often the case that a sensor-driven system captures more than just the phenomena it was designed to do so. Understanding what \emph{else} the animal data captured within an IIS may reveal, whether directly, or by processing it further, is vital to understand both the potential added value of such data, and how sensitive it may be--and thus, to what extent security and privacy considerations need to be given serious attention while designing the IIS.

Data captured by smart farming solutions, for example, might be commercially sensitive and pose a threat to the viability of an agribusiness if it were to leak~\cite{gupta2020security}, although such considerations are also dependent on the socio-cultural makeup of the sector and to what extent data is freely shared among colleagues and competitors~\cite{dirkagritech}. Perhaps more pressing as a challenge to the design and use of an IIS are the unexpected things that may be inferred from data. It has been argued that there ought to be a right to reasonable inferences~\cite{wachter2019right} made from personal data. Animal data is just as critical here, as for example, dog activity data has been argued to be sensitive as it both reveals information about their owners, as well as their caregiving~\cite{dirkaci}. As companion animals like dogs often have close relationships to their owners, to the extent that their activity and geo-location patterns may predict each other (e.g., making it trivial to derive when a dog owner typically leaves their house), but that the very nature of the dog-owner dyad may also mean inferred fitness of a dog may partially reveal fitness of their owner. Dog activity data, in turn, may thus become valuable for insurance companies who could use such information to optimize insurance premiums---not an unlikely future, as veterinary health groups have incorporated dog activity trackers~\cite{marspetcare}, and vendors have published white papers describing their envisioned architecture to aggregate and share pet health data with third parties~\cite{petcommunity}. The rise of stricter data protection legislation acts around the world (e.g., the European Union's GDPR, or the State of California's CCPA) thus make it ever more important to consider these potential privacy challenges upfront, and indeed, by design{~\cite{cavoukian2009privacy}}.

Data privacy concerns are shared by most data-driven systems. What makes the challenges discussed here unique to IIS engineering is their emergence out of the complex relationship between human and animal (e.g., the dog-human dyad and the behaviors that set it apart~\cite{mcgreevy2012overview}), as such data can not be trivially broken down to reflecting only human or animal actors, and requires in-depth ethological knowledge of the human-animal behaviors that lead to novel privacy challenges.
Perhaps because of this complexity, consumers seem to express little privacy concerns towards pet wearables or animal technology in general~\cite{van2020pets}. This may grow, though, as consumers tend to make such considerations only \emph{after} having purchased something~\cite{Emami-Naeini2019}. The growing consumer awareness on the risk of data breaches~\cite{karunakaran2018data} and the value of their data, which continues to increase in significance~\cite{anton2010internet}, thus make it important that IIS designers anticipate whether any sensitive data will be inadvertently captured.

Some examples of questions that should be raised during the engineering and research into a novel IIS to address these matters thus include:
\begin{itemize}
    \item How can animal data hold commercially sensitive information about the environment in which the IIS operates?
    \item How can animal data (e.g., activity, health, or location data) reveal information about the humans they interact with, both within and outside of the IIS?
    \item What implications do such matters have for compliance with data protection legislation (e.g., GDPR, CCPA)?
    \item What levels of protection in terms of e.g., data security are required for different kinds of animal data?
    \item How can we find a balance between benefiting from unexpected additional value within data, and using such data responsibly?
\end{itemize}
To do so, will require IIS researchers to involve data analytics, and engage with other fields such as e.g., data privacy and management science in order to pro-actively assess what else may be captured by the technology used within the IIS, and what risks this may pose towards the consumer, as well as the designer and vendor of the system.

\subsection{Transforming data into suggested interspecies interventions}
The overarching question for this challenge is \emph{how does the animal data an IIS processes lead to behaviors in the real world?}

The immediate value of animal data to the IIS is to aid human actors in better understanding and responding to animal actors as a result of their interaction with the monitoring technology. This means we need to carefully for processed information and suggested interventions:

\begin{itemize}
    \item \emph{How can they achieve concrete impacts?} Both the descriptive data generated by the IIS to aid in a human actor's understanding of the animals, as well as the suggestions it makes how to perform interspecies interventions need to be clear on what they will achieve, and how they will do so. As the IIS is meant to support humans in understanding and taking action, they should not require advanced animal behavioral knowledge. A dog owner should be told simple actions to take, just as a farm operator should be told what to do in the context of their experience.
    \item \emph{How can they be motivating and persuasive?} Suggested interventions are meant to intervene in the behavior or patterns of an animal based on objective foundations. They should thus be persuasive to the point that human actors will not second guess them or only partially execute them. At the same time, the impact of interventions may only arise on the long-term, through the result of accumulated interventions serving external processes. The IIS should thus actively motivate users if interspecies interventions do not readily have clear impacts, so that long-term benefits are not lost due to lack of action.
\end{itemize}

\subsubsection{How can they achieve concrete impacts?}
For suggested interventions to be actually performed, let alone well, they need to be understandable for their users. Such users may more more frequently laymen than not when it comes to animal behavior and physiology. This covers two important parts of the information processed by the information technology within an IIS: first the descriptive information, showing what the monitoring technology has measured and inferred, and second, the instructive information, suggesting what actions ought to be taken on basis of that information.

Consider a typical dog activity tracker. Visualizing a step count is a simply, informative way to inform the human users of the dog's activity. But doing so without contextualizing this data in what is normal for a given breed of dog will invariably lead to misinterpretation with potentially dangerous side-effects of over- or under-training as users try to correct activity based on their own human context. Other challenges for generating descriptive information come with scale. Consider, in the farm animal domain, a given farm may have hundreds of cows all being monitored by the IIS. Simply showing sensor-data of all cows at the same time is not efficient, but transforming it into aggregate data or deciding how to bring individual cows to a user's attention based on deviations in sensor values are far from trivial as well.
Thus, significant effort needs to be taken to ensure that information generated by the IIS is cognitively effective and appropriate to the types and scale of animal(s) monitored by the IIS, that is, that the information can be readily comprehended by its users without conferring unintended additional information.

But an IIS does more than just describe. It suggests interspecies interventions on basis of the data it measures. These suggested interventions, equally, need to be made as simple as possible to maximize chances that a user can carry out it well and achieve the intended impact. A suggested interspecies intervention may require more than just a technology-mediated act (e.g., changing values for feed release or temperature sensors in a cow shed). Rather, it may involve direct interaction with an animal (e.g., exercising a dog or trimming a cow's hoofs) or the animal's environment (e.g., ensuring there are items to play and cognitively enrich with, evaluating walking surfaces to prevent injury). It may even involve suggestions for others to interact with the animal, like suggesting veterinary checkups or care. Thus, suggestions where users directly interact with an animal or its environment should be carefully designed in collaboration with experts from veterinary science and animal behavior, effectively guiding users through the act and instructing them what to do, and how to expect the animal to potentially react to it. Indeed, in order to ensure beneficial impact, such suggestions need to be more instructive than simply `perform this act', but need to anticipate potential complications and instruct users when to withdraw from doing it (e.g., when a dog exhibits aggression to suggested interactions, or changes in farm animals' environment leads to unexpected behavior from individual cows or the cattle).

Some examples of questions that should be raised during the engineering and research into a novel IIS to address these matters thus include:
\begin{itemize}
    \item How can we determine the minimal intervention to achieve a given desired impact?
    \item How can we design the most cognitively effective means to convey animal behavior and physiology to people with different levels of understanding of such matters?
    \item How can animal data be related to human actor's experiences and worldviews?
\end{itemize}
This will require collaboration with veterinary science in order to understand what minimal actions might lead to concrete impact, as well as to determine what information is most salient to present to users, further use of data analytics to do so at scale for systems involving large numbers of animal actors, and importantly, consideration of cognitive science to understand how to best convey the intended information to the human actors who have to perform these interventions.

\subsubsection{How can they be motivating and persuasive?}
Descriptive information being understandable and concrete is an important first step. But the instructive information needs also to be persuasive. Suggestions \emph{grounded in} veterinary science and animal behavior expert advice, and tailored to the individual animal, need to motivate and persuade users of the IIS to perform them. 

For users to perform suggestions, it is important for the IIS and its suggestions to be perceived as useful and easy to perform~\cite{chan2009roles}. Thus, suggestions need convince users that they will be useful to themselves and the animals they interact with, whether on the short-term (e.g., if you play with your dog its stress levels may go down~\cite{coppola2006human}, if you reduce the cow shed's ambient temperature now, the cows will experience less heat stress), or on the long-term. Moreover, usefulness may be further framed in context of the user (e.g., if you play with your dog, \emph{your} stress may go down~\cite{pendry2019animal}, if you reduce the cow shed's ambient temperature now, quality of produced milk may go up~\cite{cowheat}). Such suggestive power is even more important when impact only arises on the long-term through the processes that interspecies interventions serve, whether by long-term health effects from increased activity and diet for a pet dog, or optimization of feed strategies for dairy cattle on operating costs. Thus, the content and tone of suggestions need to be carefully considered with both veterinary science and animal behavior experts and the real-world (business) context in which the IIS is situated to determine how to best convey mutually positive benefits for both human and animal.

Some examples of questions that should be raised during the engineering and research into a novel IIS to address these matters thus include:
\begin{itemize}
    \item How can we determine optimal content and format of suggested interventions for different kinds of users?
    \item How can suggested interventions also inform users accurately of what impacts to expect?
    \item How can we motivate users to perform interspecies interventions for long-term benefit?
\end{itemize}
This will require further collaboration with veterinary science and animal behavior experts, but also incorporate expertise from behavioral science to understand how motivational theories can inform persuasive interventions.

\subsection{Understanding the impact of interspecies interventions}\label{sec:understandingimpact}
The overarching question for this challenge is \emph{how do interspecies interventions affect human and animal actors?}

The interspecies interventions performed by human actors have concrete impacts on the other actors in the IIS. Such impact may be harmful or beneficial to different actors, and the impact itself may only materialize on the long-term. This means we need to carefully consider several points:
\begin{enumerate}
    \item \emph{What symbiosis is realized by impacts?} The benefits of interspecies intervention may materialize for human actors (e.g., reduced operating costs in a farm animal context), animal actors (e.g., a pet's loss of excess weight), or for both human and animal actors alike (e.g., increased health for a dog and owner from an increase in shared physical activities). Ideally, an IIS enables interventions which benefit all actors involved, regardless of their species. It is important to understand and anticipate what interventions lead to benefits for whom.
    \item \emph{Can bi-directional interspecies impact be realized?} It is typically the human actor who performs an intentional interspecies intervention suggested by the IIS, and in doing so has an impact on the animal actors. Yet, impact from actions taken by animal actors towards human actors seems to arise as well, and may become part of intentional design.
\end{enumerate}

\subsubsection{What symbiosis is realized by impacts?}
The impact of the interspecies interventions and the processes they serve may be harmful or beneficial for actors of different species within the IIS. Ideally, no IIS would suggest interspecies interventions that are actively harmful for an actor of any species (e.g., avoiding amensalistic or parasitic impacts). But understanding whether the immediately obvious short-term impacts of interspecies interventions on animal actors are beneficial or harmful is a complicated matter, which requires collaboration with veterinary and animal behavior experts. Similarly, understanding whether short-term impacts may be beneficial for business or organizational goals that the IIS contributes to (e.g., operating costs of a farm system) requires collaboration with business and management science experts through e.g., enterprise modeling efforts to predict long-term benefits of repeated short-term interventions. 

An IIS, ideally, will allow its users to reflect on the impacts of the suggested interventions and reinforce those with positive benefits, while avoiding those with negative benefits (unless perhaps they are considered to lead to more important long-term benefits by veterinary experts). Through doing so, the IIS as a socio-technical system, may reflect on itself and rise to become a system of mutualistic benefit for all involved species.

Mutualistic impacts are beneficial not just to the (different species of) users within the IIS, but to its developers as well. For example, systems which consistently achieve mutualistic impacts may allow for additional marketing value of the IIS, as some dog activity trackers already do by emphasizing the joint human-dog benefits for physical and mental health their technology provide.

Some examples of questions that should be raised during the engineering and research into a novel IIS to address these matters thus include:
\begin{itemize}
    \item How can we anticipate whom an interspecies intervention will be beneficial for?
    \item How can we clearly distinguish between symbiotic levels of impact (i.e., amensalistic, parasitic, commensalistic, mutualistic) of interspecies interventions?
    \item How can we anticipate whether desired impacts (e.g., optimiziation of cost) may be antagonistic between species?
    \item How can impacts of interspecies interventions be best observed and relayed to actors?
    \item How can we design interspecies interventions to promote mutualistic impacts?
\end{itemize}
Doing so will require extensive analysis and design research from an IS point of view, working together with veterinary science and animal behavior experts to understand how impacts on animals ought to be understood, and with experts in behavioral and management science to understand how impacts on people and business ought to be understood.

\subsubsection{Can bi-directional interspecies impact be realized?}
Most of this article has focused on human actors realizing interspecies interventions towards animal actors. But, as some of the examples discussed in Section~\ref{sec:examples} noted, animal actors may also interact with the technology in the IIS to communicate with human actors and motivate them to act. The example of the audio/video-enabled feeding system where a dog could bark at the device, prompting their owners to release more food from a distance presents the possibility of animals learning how to interact with technology to motivate human actors to do things. This is, of course, not a data-driven intervention in the same vein as human actors performing interspecies interventions.

Yet, the potential of studying if and how animal actors manage to use the IIS directly to address their own needs may allow for richer behavioral data captured by monitoring technology which, in turn, can enrich the suggested interspecies interventions by adding an understanding of how the animal actor may react to certain interventions or the lack thereof.

Some examples of questions that should be raised during the engineering and research into a novel IIS to address these matters thus include:
\begin{itemize}
    \item How do we know when and whether animal actors can use the IIS to actively communicate needs to human users?
    \item Can the socio-technical structure of an IIS be designed to actively reflect and react on the behavior of its animal actors?
\end{itemize}
Doing so will require extensive analysis of deployed IIS, working with data analytics experts to understand whether patterns of animals using the IIS to communicate needs towards humans have already occurred, and how they may be characterized. Moreover, if such patterns can be identified, the design of an IIS requires further collaboration with veterinary science and animal behavior experts to understand how to design the technology and actions people take in order to allow animal actors to be `equal citizens' within the IIS.

\section{Concluding outlook---the benefits and future of tackling interspecies information systems}\label{sec:conclusion}
Interspecies information systems bring many challenges in their engineering and use. Some of these challenges can be tackled in isolation as research into requirements for technology for animals has done so far: ensuring hardware fits with animal physiology, ensuring that technology is usable for all actors.
But new data-driven challenges emerge when we take an information systems perspective and analyze how animal data flow throughout an IIS and informs human behavior towards animals.
To that end, this paper offers two key contributions to aid in the practical (re-)design of technology for animals.
\begin{enumerate}
\item It provides a a detailed analysis of the typical elements and properties of an IIS culminating in a conceptual model (Section{~\ref{sec:model}}) and its related detailed explanations (Sections{~\ref{sec:commonalities}}--{~\ref{sec:interspeciesimpact}}). This supports designers in analyzing not only what kind of IIS has, or will emerge through use of the technology for animals they design, but also in looking beyond the technology itself and considering its wider environment and context. Doing so then opens up considerations of ongoing challenges specific to IIS from the very first design phases.
\item It sets out in detail those ongoing challenges (Section{~\ref{sec:challenges}}) that will typically be faced in designing IIS, providing clear questions that should be explored, and who to do so with. This supports designers in tackling challenges from understanding the potential of animal data, transforming it into suggested interspecies interventions, and finally understanding the impact of those interventions \emph{by design}, rather than having to address them after they arose.
\end{enumerate}

This article has shown that designing technology which enables an IIS need to account for several challenges: understanding the potential of animal data, transforming it into suggested interspecies interventions, and understanding the impact of those interventions. Tackling these challenges requires an approach that involves extensive inter-disciplinary work, collaborating with experts across fields that deepen our understanding of animals, the data they produce, and how this affects the real world. 

In looking to the future of designing and engineering IIS, the decidedly anthropocentric focus of the examples analyzed here needs to be mentioned. Such an anthropocentric focus may be damaging to eventual goals of (at least some types of) IIS, by reaffirming the sole pursuit of human interests, possibly damaging animal lives{~\cite{evans2019there}}, and may ironically reinforce human sense of dominion over animals from the information asymmetry resulting out of their use{~\cite{kamphof2013linking}}. Thus, especially those IIS seemingly in the pursuit of joint human and animal best interests--such as improving caregiving of companion animals, or conserving wildlife species, should endeavour to consider how (if at all possible) this information asymmetry can be meaningfully resolved in the future. Examples of research being conducted that may in the future lead to such work is that of designing technology that informs humans of how animals conceive the world by embracing their otherness{~\cite{chloe2020}}.

The promise that animal-centered technology bring for our ability to ensure animal welfare, improve our care-giving or management of animals (even if for decidedly human interests in some cases), and strengthen our our joint health and well-being cannot be denied. It should be imperative that we stimulate the use of animal-centered technology, but in doing so, carefully and systematically assess the challenges that the IIS that emerge through their use bring.

%% file: table1.tex
\renewcommand{\arraystretch}{1.2}
\begin{table*}[htb!]
\caption{Definitions of elements and relationships in the IIS model, explained with partial example instantiations of dog activity tracking and cattle management.}
\centering
\small
\begin{tabularx}{\textwidth}{p{1.7cm}|p{2.5cm}|p{4cm}|p{4cm}}

\hline \textbf{Element}	&	\textbf{Definition}	& \textbf{Example: Dog activity tracking}	& \textbf{Example: Cattle management} \\ \hline

\multicolumn{4}{c}{\emph{Entities}} \\ \hline

\cellcolor{yellow!40} Human actor   &   human beings  & Fede the dog owner & Dmitry the dairy farmer \\
\cellcolor{yellow!40} Animal actor   &   non-human beings  &  Nunzio the Volpino Italiano & A herd of dairy cattle \\

\cellcolor{green!40} Monitoring technology &   technology capturing data  & FitWoof collar & Worn accelerometer and vital sign sensors, temperature sensor in cow shed \\
\cellcolor{green!40} Information Technology    &   technology processing data  & Fede's smartphone running the FitWoof app & Dmitry's computer running CowManager 2.0\\

\cellcolor{cyan!30} Raw Data  &   data captured by monitoring technology  &   3-axis accelerometer data showing average body acceleration on 1 minute intervals & Raw data of time spent in feeding area, weight gain over time, vital signs, environment temperature \\
\cellcolor{cyan!30} Processed Data    & data processed by information technology    & Nunzio's daily stepcounts, recommended activity levels & Assessment of caloric intake and expenditure by activity, acceptable temperature range for breed \\
\cellcolor{cyan!30} Policy    & suggested interspecies interventions informed by processed data    & Volpino Italiano dogs should be walked for at least 60 minutes per day. & Dairy cows should not be exposed to ambient temperatures higher than 30$^{\circ}$C \\
\cellcolor{violet!40} Interspecies intervention & an act realized by human actors affecting non-human actors    & Fede taking Nunzio for a 30 minute walk & Reducing feed to account for caloric surplus, decrease environmental temperature to increase activity \\

\cellcolor{violet!40} Process   &   an external process affected by interspecies interventions    &   Increasing activity levels appropriate to breed & Optimization of feed for breed and environmental conditions to reduce operating costs \\ \hline

\multicolumn{4}{c}{\emph{Relationships}} \\ \hline

serves  & $x$ is input to $y$   &   Nunzio is input to FitWoof collar & Cattle data is input to CowManager 2.0\\
uses  & $x$ interacts with $y$   &   Fede uses FitWoof app to see how active Nunzio has been today & Dmitry uses CowManager 2.0 to assess optimization points  \\
suggests  & $x$ suggests to perform $y$   &   FitWoof app shows Nunzio has been inactive most of the day and based on prior activity patterns suggests to take Nunzio for a 30 minute walk & CowManager 2.0 suggests Dmitry to reduce feed release and turn on air-conditioning \\
realizes  & $x$ realizes $y$   &   Fede takes Nunzio for a 30 minute walk & Dmitry reduces feed release and turns on air-conditioning \\

impacts  & $x$ has an impact \{harmful, beneficial\} on $y$   &   Sustained frequent activity has a beneficial impact on Nunzio's physical fitness & Turning on air-conditioning has a beneficial impact on cattle welfare; Optimization of feed has a beneficial impact on Dmitry's operating costs \\

\hline 
\end{tabularx}
\label{tab:definitions}
\end{table*}
\renewcommand{\arraystretch}{1}